\documentclass[a4paper,11pt]{article}
\usepackage{pos}

\title{Search for dark matter from the center of the Earth with 8 years of IceCube data}

\author{The IceCube Collaboration \\{\normalsize \normalfont(a complete list of authors can be found at the end of the proceedings)}}

\emailAdd{grenzi@icecube.wisc.edu}
\emailAdd{grenzi@ulb.be}

\abstract{The nature of Dark Matter (DM) remains one of the most important unresolved questions of fundamental physics. Many models, including Weakly Interacting Massive Particles (WIMPs), assume DM to be a particle and predict a weak coupling with Standard Model matter. If DM particles can scatter off nuclei in the vicinity of a massive object such as a star or a planet, they may lose kinetic energy and become gravitationally trapped in the center of such objects, including Earth. As DM accumulates in the center of the Earth, self-annihilation of WIMPs into Standard Model particles can result in an excess of neutrinos which are detectable at the IceCube Neutrino Observatory, situated at the geographic South Pole. A search for excess neutrinos from these annihilations has been performed using 8 years of IceCube data, and results have been interpreted in the context of a number of WIMP annihilation channels ($\chi\chi\rightarrow\tau^+\tau^-$/$W^+W^-$/$b\bar{b}$) and masses ranging from 10 GeV to 10 TeV. We present the latest results from this analysis and compare the outcome with previous analyses by IceCube and other experiments, showing competitive results, which are even world-leading in some parts of the parameter space.

\vspace{4mm}
{\bfseries Corresponding authors:}
Giovanni Renzi$^{1*}$\\
{$^{1}$ \itshape Université Libre de Bruxelles}\\
$^*$ Presenter
}
\FullConference{37$^{\rm{th}}$ International Cosmic Ray Conference (ICRC 2021)\\
		July 12th -- 23rd, 2021\\
		Online -- Berlin, Germany}

\begin{document}
\maketitle

\section{Introduction}
\label{sec:intro}
Astronomical  evidence for the existence of Dark Matter (DM) has been observed throughout the whole last century. The nature of DM remains unknown and several models have been proposed to describe it
\cite{Bertone:2005}. Among others, particle DM models stand out for the number of experiments that have been
probing them. The Weakly Interactive Massive Particle (WIMP) is one of the most discussed candidates and is
predicted by supersymmetric extensions of the Standard Model (SM). Such particles could scatter off SM particles in
the vicinity of a massive celestial body, such as the Sun or the Earth, lose energy and be gravitationally trapped
in the center of the body. WIMPs accumulated in this way can self-annihilate into SM particles with a rate that is
proportional to the DM density in the center of the body. Neutrinos can be produced in this process and their
particular properties allow them to travel through dense environments and for long distances. For the Earth, they
can reach the surface of the planet and potentially be detected.

The IceCube Neutrino Observatory can measure the flux of neutrinos coming from the center of the Earth. Previously, a search with 1 year of data \cite{EarthIce:2016} has produced upper limits on the spin-independent WIMP-nucleon
cross-section. in these proceedings the latest developments on the new search using 8 years of data will be
presented. 

\section{The IceCube detector}
\label{sec:ic}
IceCube \cite{IceCube:2016a} is a cubic kilometer neutrino detector located at the geographic South Pole and
installed in the Antarctic ice between depths of 1450 m and 2450 m. The detector consists of a large
array of photomultipliers (PMTs) housed in glass spheres called Digital Optical Modules (DOMs).
IceCube is composed of 86 vertical strings with 60 DOMs each and vertical spacing of 125 m. The
DOMs record Cherenkov light emitted along the path of relativistic charged particles produced by
neutrino interactions. The collected light allows reconstruction of the characteristics of the primary
neutrinos such as energy and direction. Inside the IceCube volume, a smaller and denser array at a
depth of 1750 cm, called DeepCore, is also installed. It consists of 8 closely-spaced strings in the
center of the primary array with average sensor spacing of 72 m. DeepCore can use the remaining
instrumented volume as a veto against muon and neutrino events originating from atmospheric
interactions. DeepCore is of particular importance to detect neutrinos with energy below 100 GeV.

\section{Dark matter from the center of the Earth}
\label{sec:dm}
Dark matter particles can be captured in the center of the Earth when they lose energy by scattering off nucleons in
the planet. The relative abundance of elements in the Earth sets the capture process to be led by the
spin-independent DM-nucleon cross-section $\sigma_{\rm SI}$. The capture rate $C_{\rm C}$ depends further on the DM
particle mass and the DM density and velocity distribution in the Solar System. Self-annihilation can thereby happen
in the center of the Earth, led by the velocity averaged self-annihilation cross-section 
$\langle\sigma_{\rm A} v\rangle$. The full process is described by \cite{Bruch:2009}:
\begin{equation}
    \label{eq:dm_flux}
    \dot{N} = C_{\rm C}-C_{\rm A}N^2-C_{\rm E}N,
\end{equation}
where $\Gamma_{\rm A}=C_{\rm A}N^2$ is the annihilation rate and depends on $\sigma_{\rm A}v$ and the age of the
Earth $t_{\oplus}\simeq4.5\times10^9 \textrm{ y}$. The evaporation term $C_{\rm E}N$ can be neglected for masses
bigger than a few GeV \cite{evaporation:2021}, therefore \eqref{eq:dm_flux} has solution
\begin{equation}
    \label{eq:dm_sol}
    \Gamma_{\rm A} = \frac{C_{\rm C}}{2}\tanh^2\left(\frac{t}{\tau}\right),
\end{equation}
where $\tau=(C_{\rm C}\cdot C_{\rm A})^{-1/2}$. \eqref{eq:dm_flux} reaches equilibrium when $t\gg \tau$. The
equilibrium have not reached for the Earth case. This means that, to compute $\sigma_{\rm SI}$, a value for 
$\langle\sigma_{\rm A}v\rangle$ must be assumed.

The Standard Halo Model (SMH) \cite{DM_rev:96} is the galactic dark matter description adopted for this work. It
assumes a truncated Maxwellian velocity distribution with a dispersion of 270 km/s and an escape velocity of 544
km/s. The local DM density value is under debate \cite{Read:2014} and could vary between $\sim0.2\textrm{ GeV/cm}^3$
and $\sim0.5\textrm{ GeV/cm}^3$. For this analysis, in consistency with the other searches in the field, the value
$0.3\textrm{ GeV/cm}^3$ is assumed.
\begin{figure}
    \centering
    \includegraphics[width=.6\textwidth]{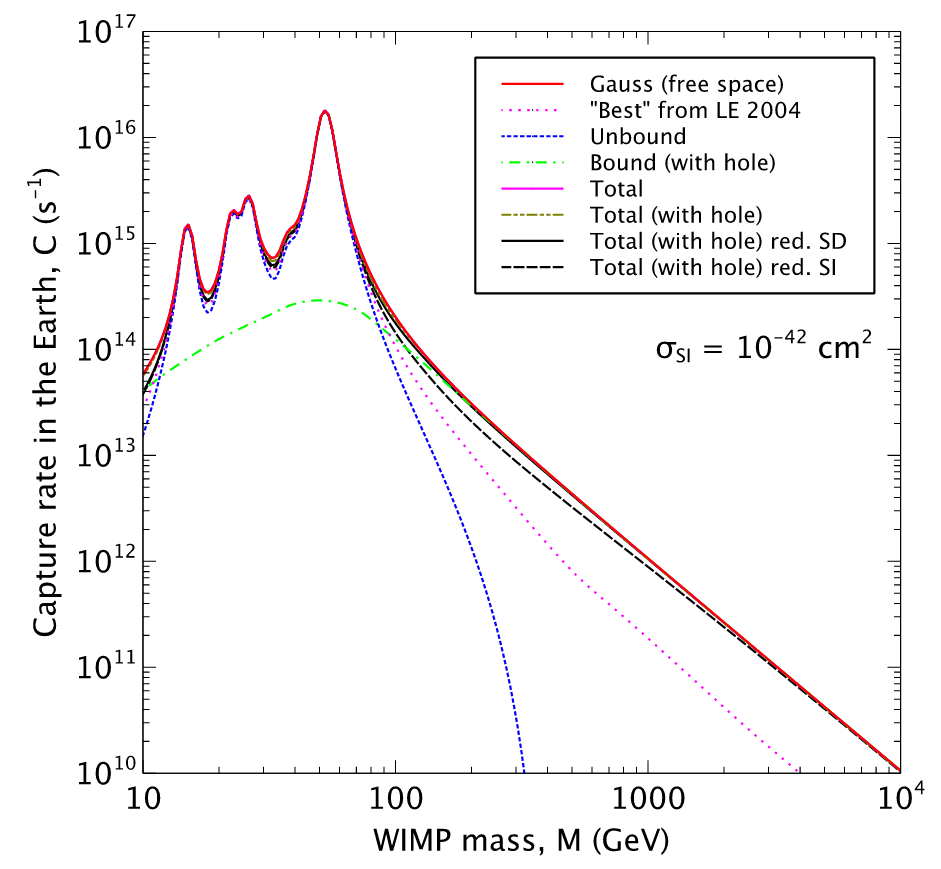}
    \caption{Capture rate as a function of the DM mass assuming $\sigma_{\rm SI}=10^{-42}\textrm{ cm}^2$.
    \cite{Sivertsson:2012}}
    \label{fig:capture}
\end{figure}

IceCube can detect the neutrino flux reaching the surface at the South Pole, which can be described by
\begin{equation}
    \label{eq:nu_flux}
    \Phi=\frac{\Gamma_{\rm A}}{4\pi R^2_{\oplus}}\frac{d N_{\nu}}{d E_{\nu}},
\end{equation}
where $R_{\oplus}$ is the Earth radius and $\frac{d N_{\nu}}{d E_{\nu}}$ is neutrino energy spectrum, which depends
on the DM particle mas and the annihilation channel.

\section{Datasets and event selection}
\label{sec:data}
The analysis is intended to use 8 years of IceCube data, from 2011 to 2018. The peculiar position of the source
does not allow us to rely on scrambled data to estimate the background, so Monte-Carlo (MC) simulations are
necessary. A burn sample of $\sim204$ days of data is used to check the reliability of the
simulations, in terms of distributions and overall rates. 

The signal for this analysis consists of nearly vertical, up-going tracks, which are produced when the signal
neutrinos interact in the instrumented ice volume. The main background for this analysis is the so-called
\textit{atmospheric muons}: these are produced in cosmic-ray air showers and cannot traverse the Earth so that from
the point of view of IceCube they can only be down-going events. Some of them are though mis-reconstructed as
up-going, simulating directions which are typical of the signal. Another relevant background is the
\textit{atmospheric neutrinos}, which are produced which are produced in the same interactions which produce
atmospheric muons and can traverse the whole Earth. These neutrinos are therefore up-going and interact in the
detector in the same way as for signal.

Signal simulations have been obtained with the WimpSim \cite{Wimpsim:2008} software package, which can produce
self-annihilation events in the center of the Earth and propagate them to the detector. In particular, datasets for
annihilation channels $\chi\chi\rightarrow\tau^+\tau^-$/$W^+W^-$/$b\bar{b}$ and for DM mass from 10 GeV to 10 TeV
have been produced. Atmospheric muons have been simulated with CORSIKA \cite{corsika:98}. Atmospheric neutrinos are
simulated with the NuGen and GENIE software package. Neutrino oscillation \cite{OscIce:2017} has to be
taken into account when simulating neutrinos with energies below 100 GeV.

\begin{figure}
    \centering
    \begin{minipage}{0.49\textwidth}
        \includegraphics[width=1.\textwidth]{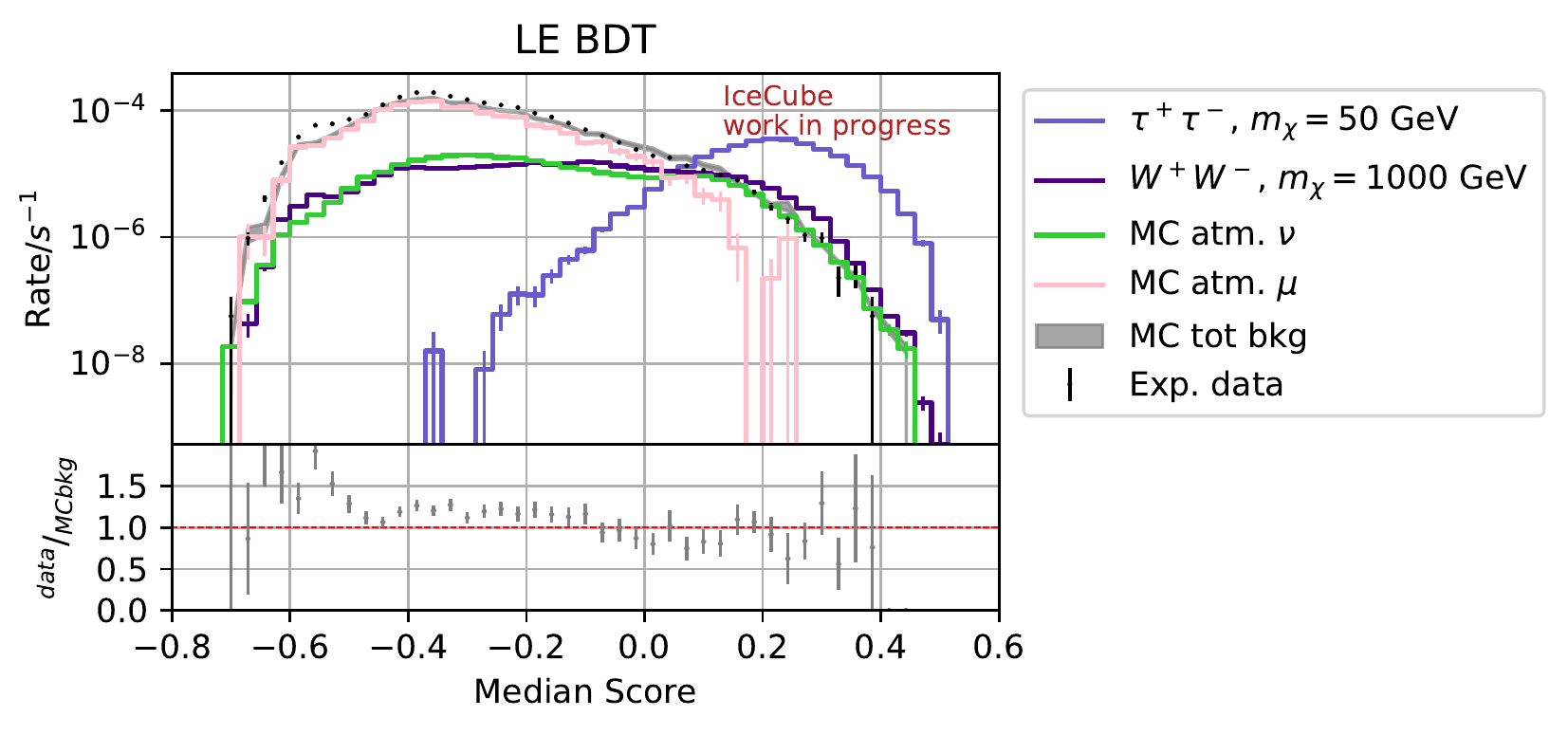}
        \caption*{Median score value distributions for the LE selection.}
    \end{minipage}
    \begin{minipage}{0.49\textwidth}
        \includegraphics[width=1.\textwidth]{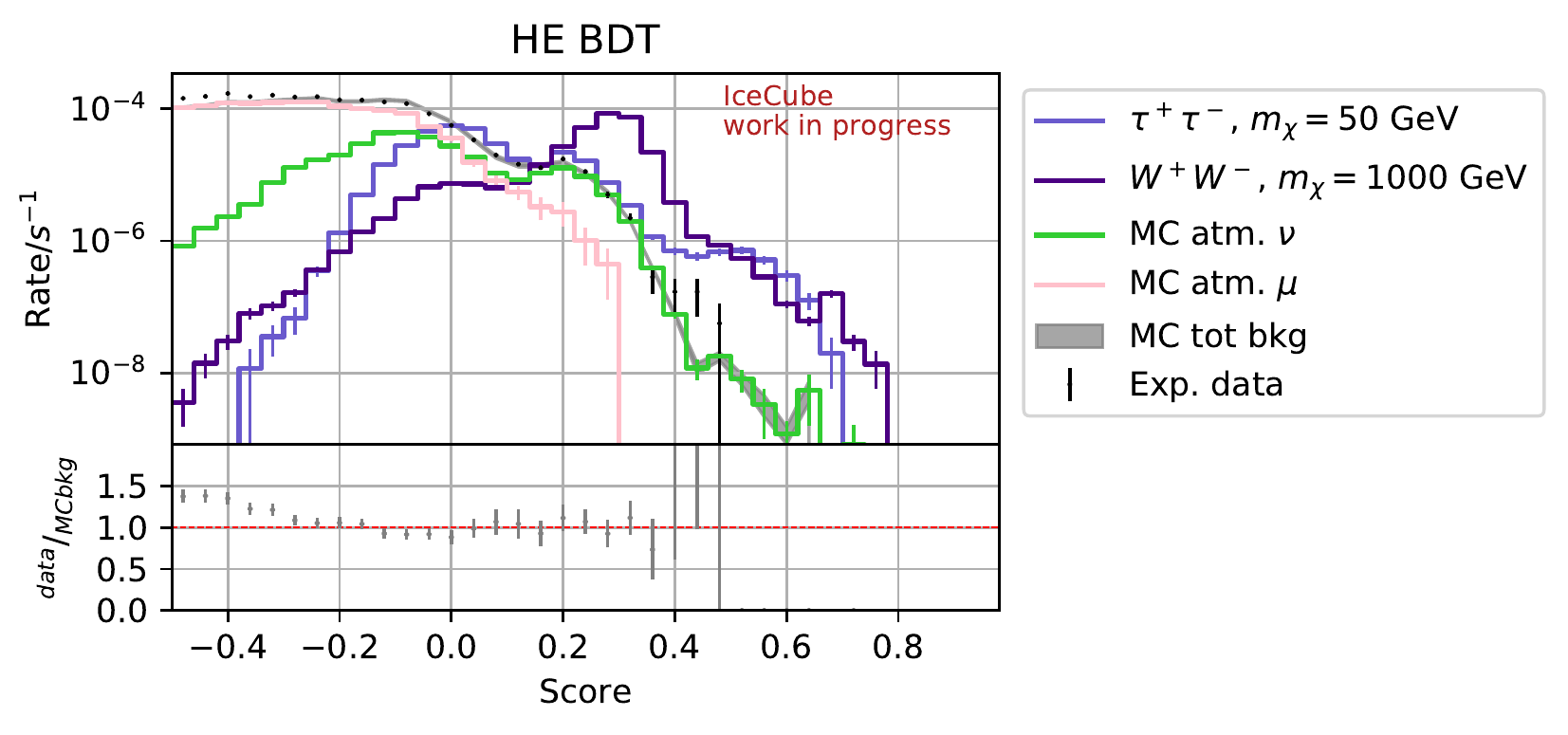}
        \caption*{Score value distributions for the HE selection.}
    \end{minipage}
    \caption{The two violet lines indicate the two baseline signal datasets: $\chi\chi\rightarrow\tau^+\tau^-$,
    $m_{\chi}=50\textrm{ GeV}$ and $\chi\chi\rightarrow W^+W^-$, $m_{\chi}=1\textrm{ TeV}$ for LE and HE
    respectively. The grey band indicates the total MC distribution. Statistical uncertainties are included.}
    \label{fig:bdt}
\end{figure}

A dedicated event selection has been developed for this analysis. The early stages of the selection are focused on
separating muons from neutrinos. This can be achieved by imposing that the interaction producing the track happens
in the detector and via some cuts on the quality of the reconstruction; the muons are mis-reconstructed, hence they
tend to have lower-quality reconstructions. The main stage of the selection is a random forest classifier. This
machine-learning tool can be trained to discriminate signal from background assigning a score from -1 (most
background like) to 1 (most signal like). The Boosted Decision Trees (BDTs) constituting the forest are fed the
burn sample data as the background set and signal simulation as the signal set. At this point, the selection is
split into High Energy (HE) and Low Energy (LE) selections, because events from the highest DM masses have been
observed to widely differ from events from the lowest DM masses. The LE selection MC statistics is, at this point,
not high enough to allow a proper treatment for the latest stages of the selection. Hence 200 random forests are
trained and the events are assigned a weight corresponding to the ratio of random forests that would pass a certain
cut on the random forest score. For the HE selection statistics are sufficient to continue the analysis, therefore
one only random forest is trained and a straight cut on the score is applied. The two cut values have been chosen
to be the values that obtain the best possible sensitivity for the LE and HE selections respectively. The cut
value for the LE selection is 0.24 and the weights can hence be expressed as
\begin{equation}
    w_{0.24} = \frac{n_{s>0.24}}{n_{\rm tot}},
\end{equation}
where $s$ is the score and $n_{\rm tot}=200$. 
The cut value for the HE selection is $s>0.18$. Distributions of the scores for the two selections are in figure
\ref{fig:bdt}.

The last stage of the selection consists of a hard cut on the reconstructed zenith value at $\theta_{\rm reco}>2.8
\textrm{ rad}$ and the performance of a more suitable energy reconstruction algorithm. It is worth underlining
that, at this stage, the atmospheric muon component of the background is completely removed. 

\section{Analysis method and sensitivity}
\label{sec:method}
\begin{figure}[h]
    \centering
    \begin{minipage}{0.49\textwidth}
        \includegraphics[width=1.\textwidth]{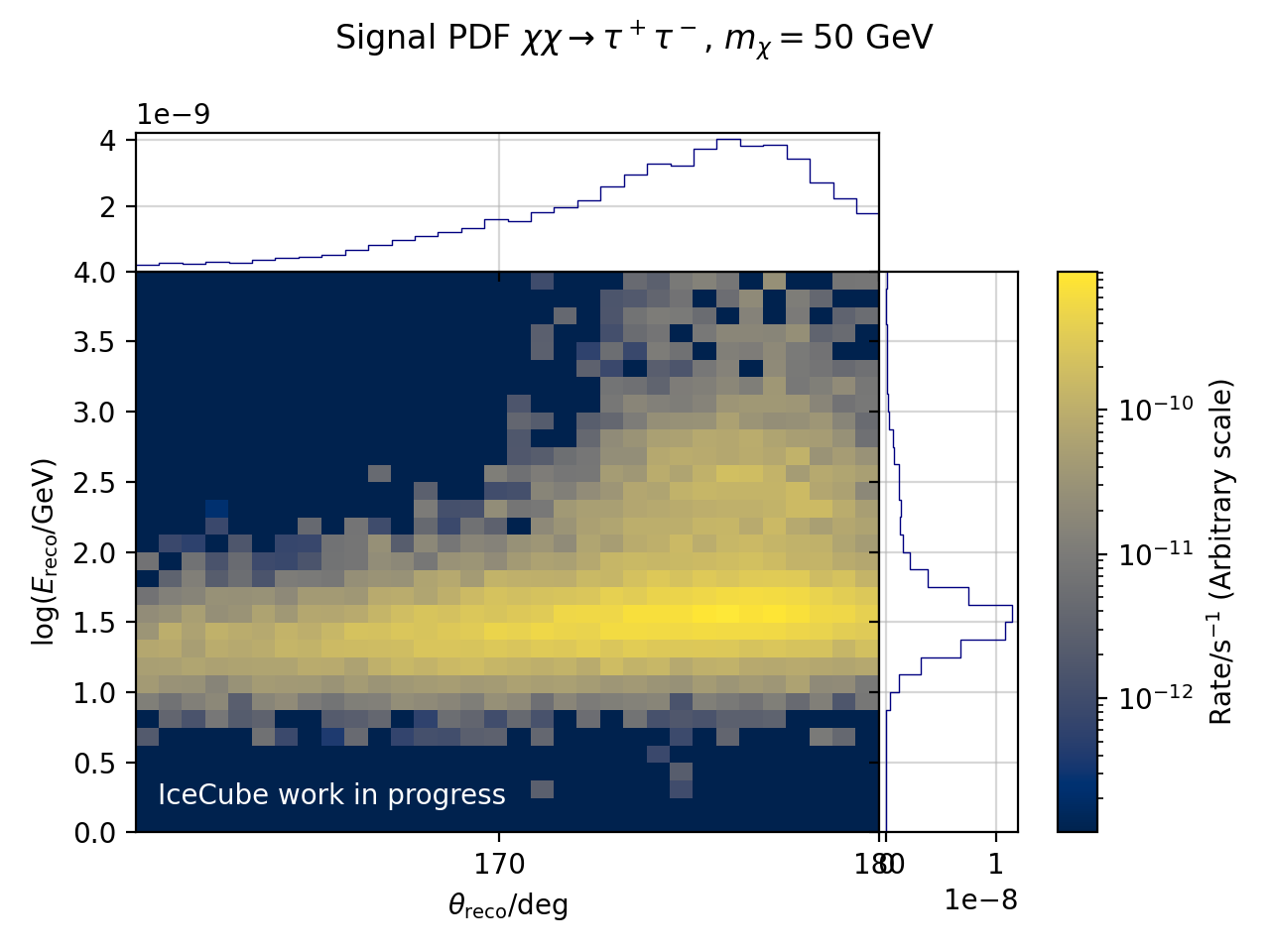}
    \end{minipage}
    \begin{minipage}{0.49\textwidth}
        \includegraphics[width=1.\textwidth]{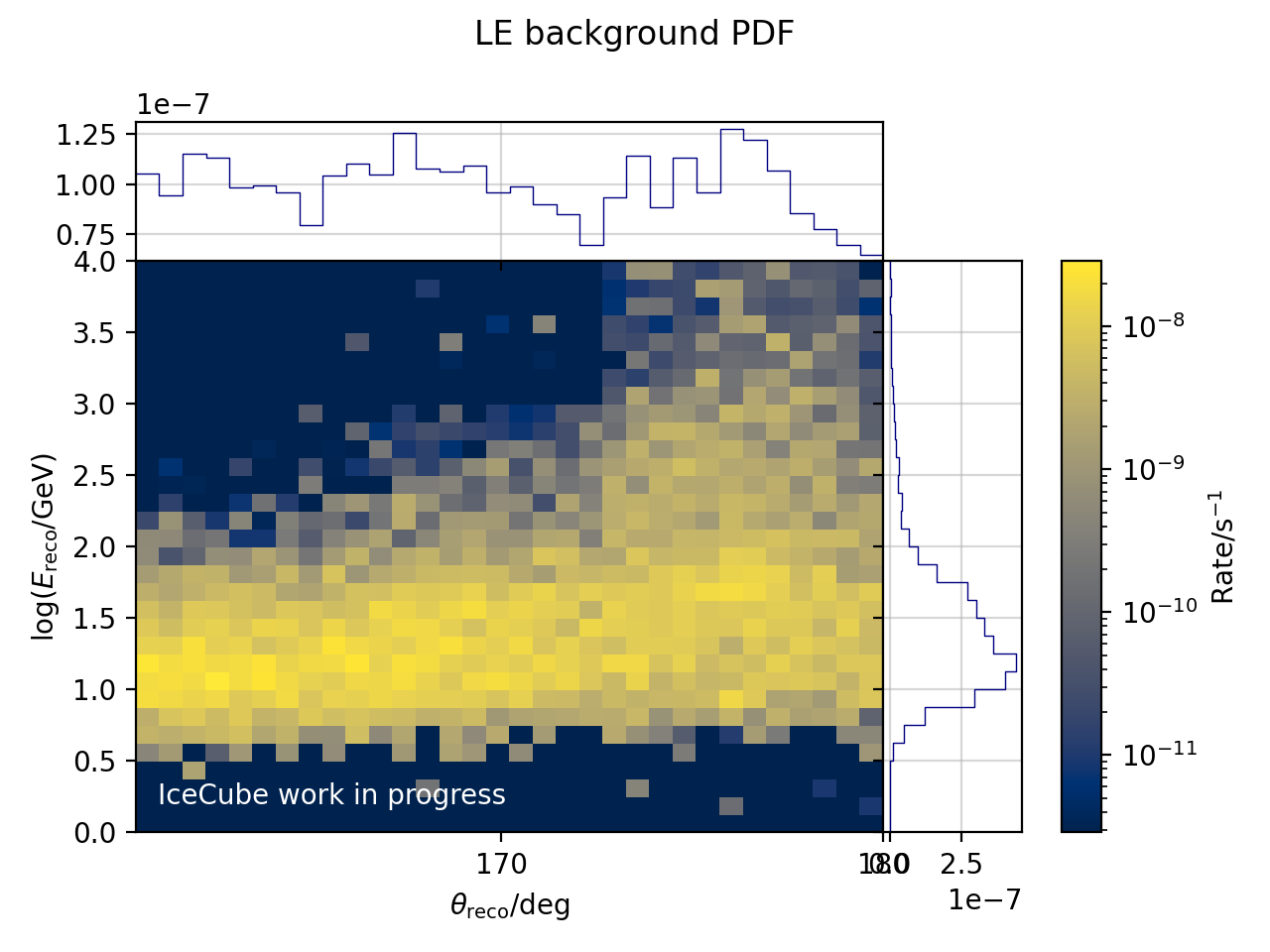}
    \end{minipage}
    \caption{Signal baseline and atmospheric background distribution for the LE selection.}
    \label{fig:le_pdf}
\end{figure}

\begin{figure}
    \centering
    \begin{minipage}{0.49\textwidth}
        \includegraphics[width=1.\textwidth]{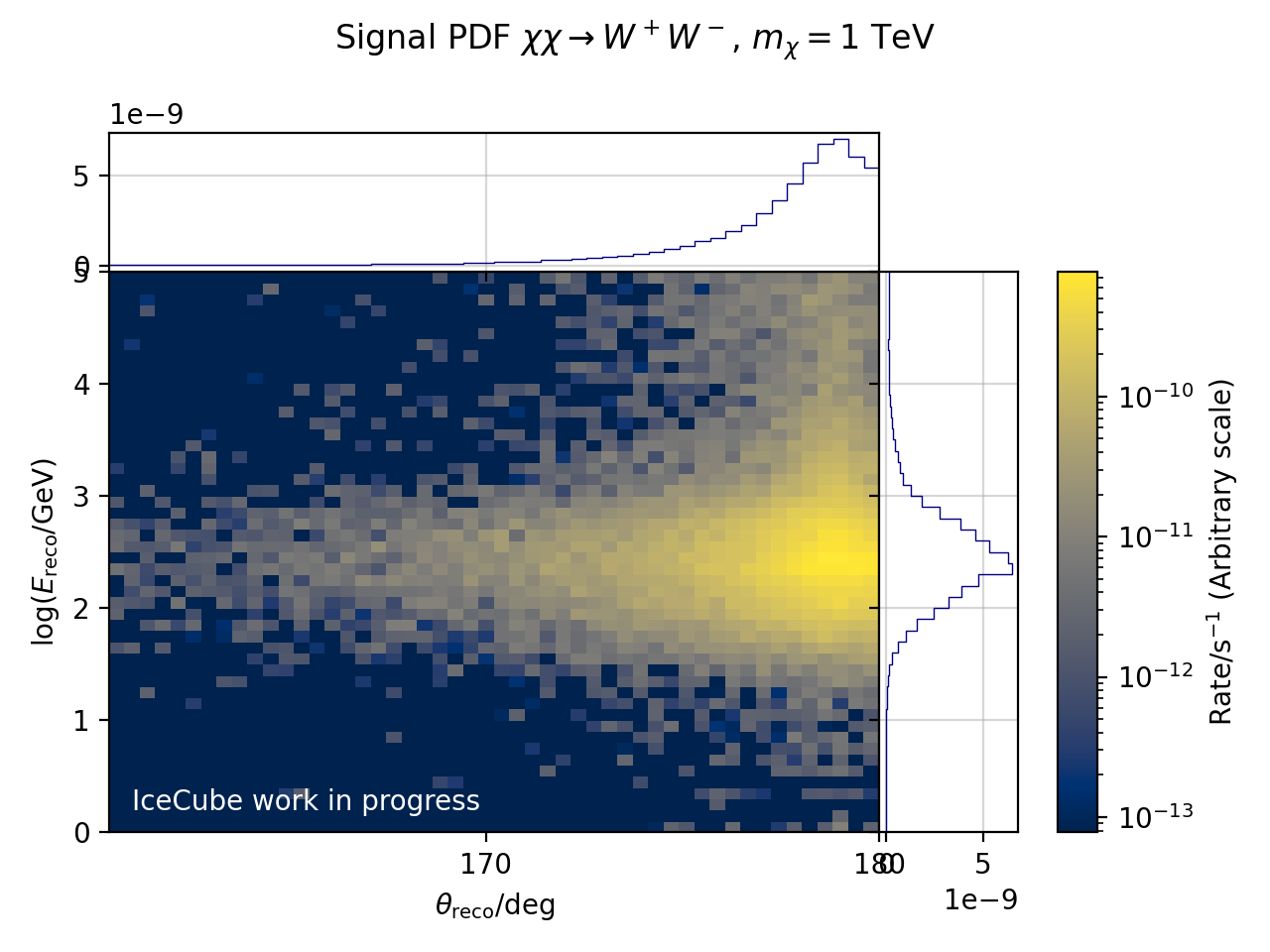}
    \end{minipage}
    \begin{minipage}{0.49\textwidth}
        \includegraphics[width=1.\textwidth]{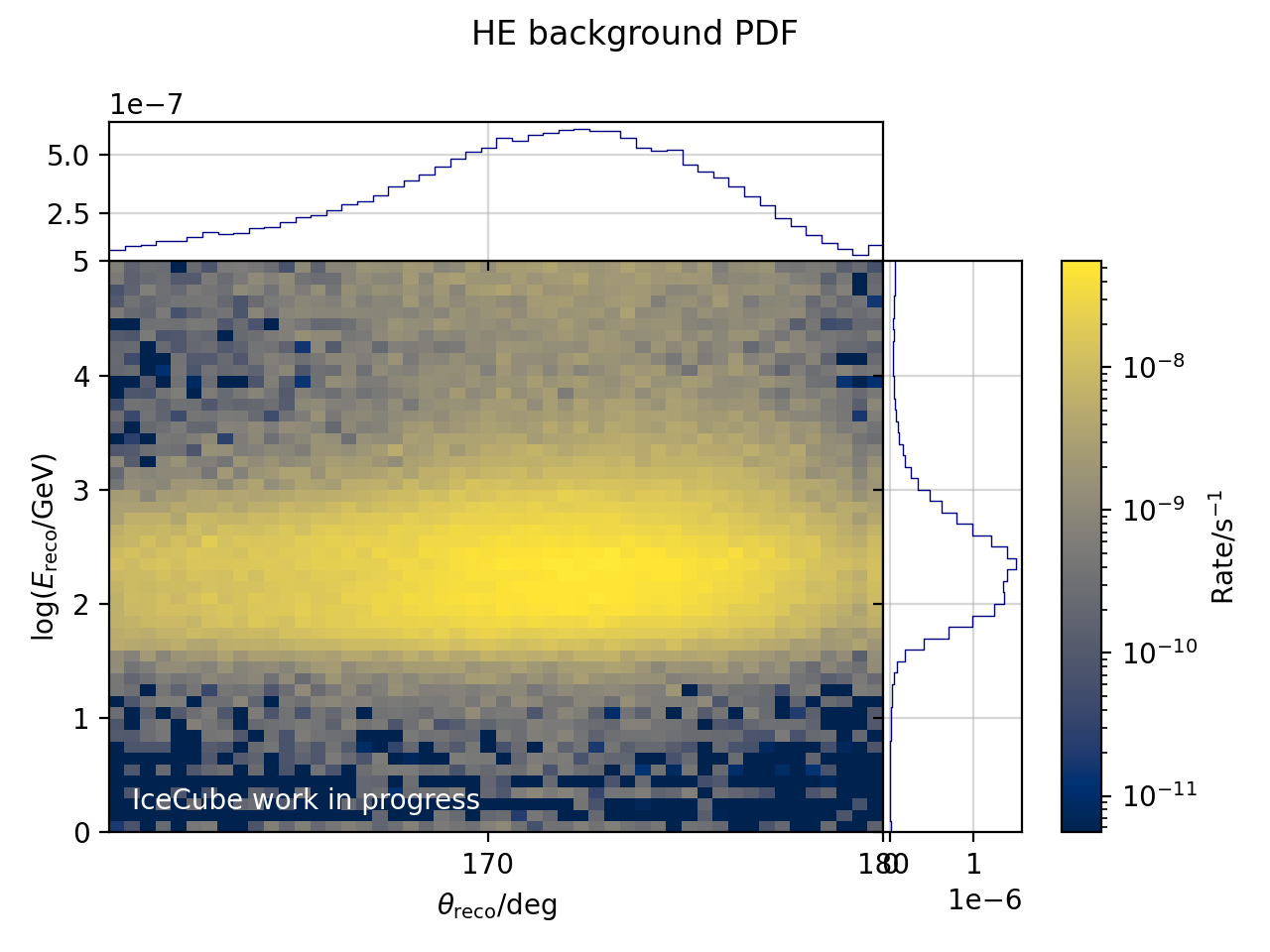}
    \end{minipage}
    \caption{Signal baseline and atmospheric background distribution for the HE selection.}
    \label{fig:he_pdf}
\end{figure}

In order to obtain limits, a binned analysis is performed. The Point Spread Function (PDF) is defined for each bin
as follows:
\begin{equation}
    \label{eq:pdf}
    \mu_{\rm bin}(\xi,\overrightarrow{\eta})=\xi S_{\rm bin}(\theta, E)+\sum_i\eta_{\rm i}B_{\rm i, bin}(\theta, E),
\end{equation}
with $N_{\rm tot}$ being the total number of events, $\xi$ is the signal fraction, $\eta_{\rm i}$ are the fractions
of the various background component, $S(\theta, E)$ is the final distribution for signal and $B_{\rm i}(\theta, E)$
are the final distributions of the background components. These are 2D zenith-energy distributions and are shown in
figures \ref{fig:le_pdf} and \ref{fig:he_pdf} for the LE and HE selections respectively. These distributions are the
for the analysis. The number of bins representing the data has been optimized to obtain the best possible
sensitivities.

The "effective" likelihood developed in \cite{Effective:2019} is used. This likelihood expression can be considered
a generalization of the Poisson likelihood, where the uncertainties on the distributions are taken into account. In
the case where the uncertainties are negligible, it reduces to the Poisson case. Given the signal fraction $\xi$ ,
 the parameters that describe the background $\overrightarrow{\eta}$, and the number of observed events $k$, the
likelihood can be expressed as
\begin{equation}
    \label{eq:llh}
    \mathcal{L}_{\rm eff}(\mu|k)=N_{\rm tot}\frac{\beta^{\alpha}\Gamma(k+\alpha)}{k!(1+\beta)^{k+\alpha}
    \Gamma(\alpha)},
\end{equation}
where $\alpha$ and $\beta$ are functions of the observed distribution $\mu$ and its uncertainty distribution
$\sigma$:
\begin{equation}
    \label{eq:alpha_beta}
    \alpha=\frac{\mu^2}{\sigma^2}+1\hspace{6 pt}\textrm{;}\hspace{6 pt}\beta=\frac{\mu}{\sigma^2},
\end{equation}
The total likelihood is the sum of the likelihoods calculated for each bin.\\
The Test Statistic (TS) is defined by the ratio of likelihood absolute best fit and the best fit obtained by
keeping $\xi=0$:
\begin{equation}
    \label{eq:ts}
    TS = \frac{\mathcal{L}(\hat{\xi},\hat{\overrightarrow{\eta}}|k)}
    {\mathcal{L}(\xi=0,\hat{\overrightarrow{\eta}}|k)}.
\end{equation}
The 90\% confidence level limit is computed via a frequentist approach. Signal events are progressively injected and
pseudo-experiments are generated at each step. A TS is computed for each trial. The limit is reached when, for a
certain number of injected signal events $n^{0.9}_{\rm inj}$, 90\% of the pseudo-experiments TS values are greater
than the median TS obtained when $n_{\rm inj}=0$. $n^{0.9}_{\rm inj}$ is hence the sensitivity at 90\% Confidence
Level (C.L.) in terms of number of events.

\begin{figure}[!h]
    \centering
    \includegraphics[width=0.82\textwidth]{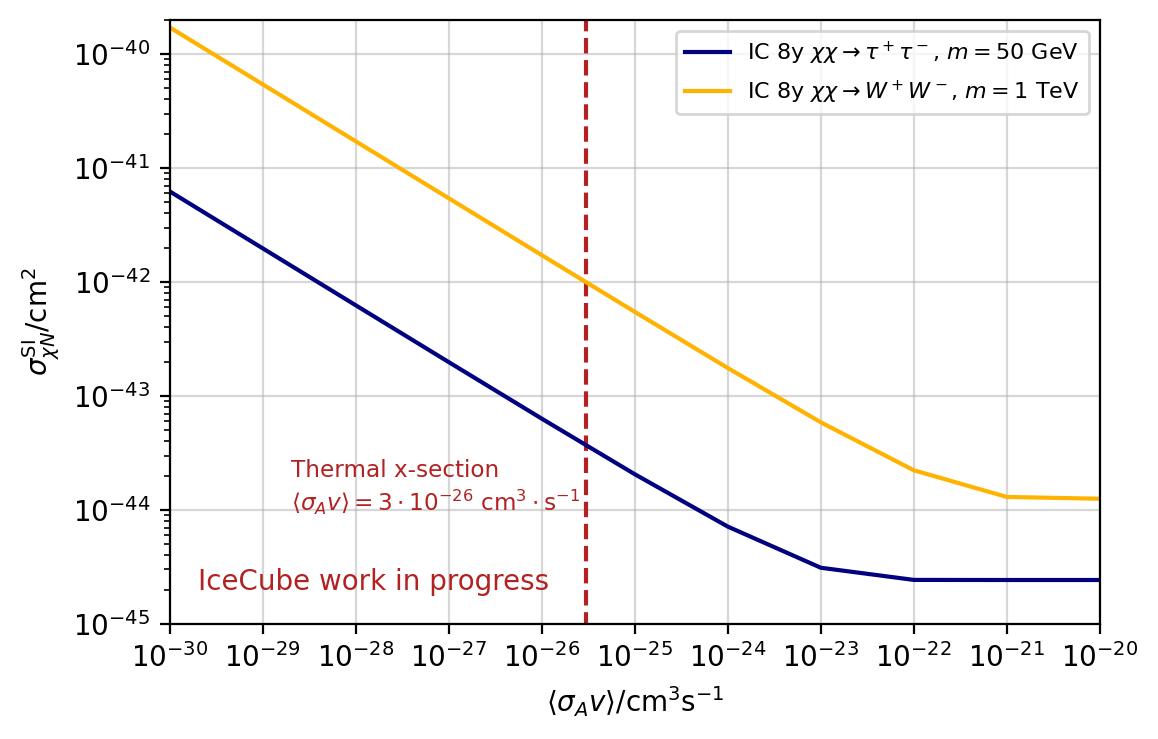}
    \caption{Sensitivity for $\sigma_{\rm SI}$ in function of the assumption made on
    $\langle\sigma_{\rm A}v\rangle$. The lines reach a horizontal plateau when the process is in equilibrium.}
    \label{fig:cross_sens}
\end{figure}

\begin{figure}[!h]
    \centering
    \includegraphics[width=0.82\textwidth]{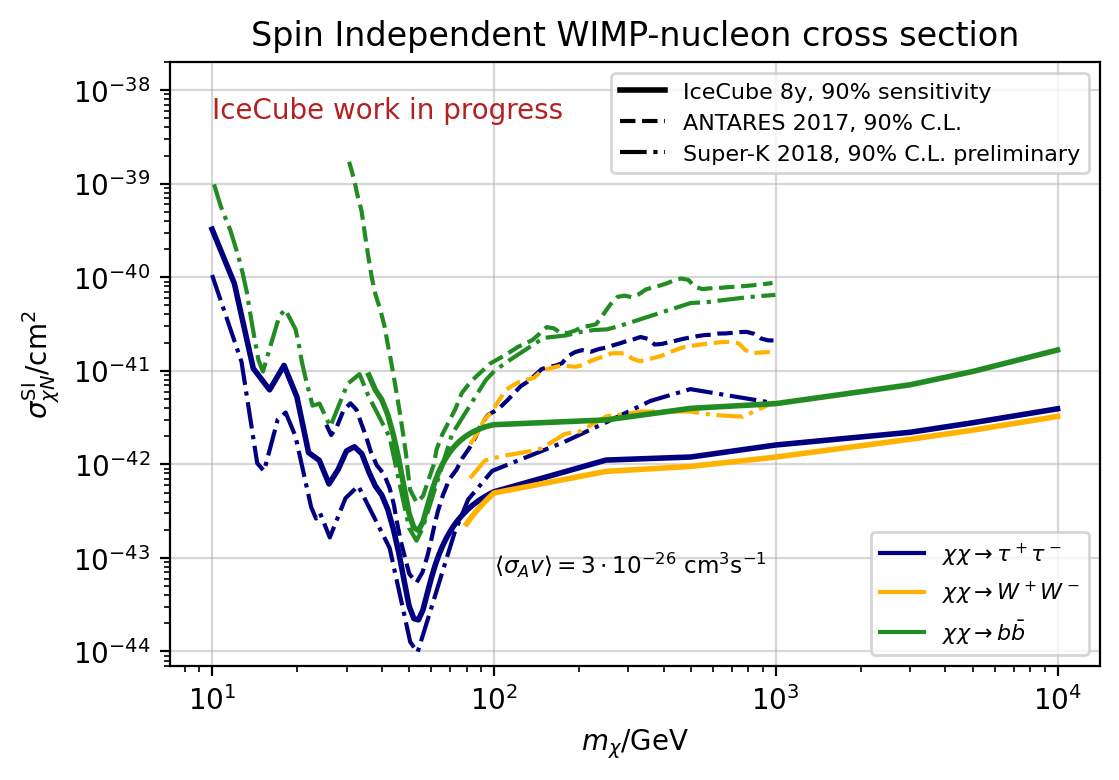}
    \caption{Sensitivities at 90\% C.L. compared to the latest results from ANTARES \cite{ANTARES_Earth:2016} and Super-Kamiokande \cite{SuperK_DM:2020}.}
    \label{fig:xs_sens}
\end{figure}

The volumetric flux limit can be computed as follows:
\begin{equation}
    \label{eq:vol_flux}
    \Gamma_{\nu\rightarrow\mu}=\frac{n^{0.9}_{\rm inj}}{t_{\rm live}V_{\rm eff}},
\end{equation}
where $t_{\rm live}$ is the livetime of the analysis and $V_{\rm eff}$ the effective volume of the detector.
The volumetric fluxes can then be converted to annihilation rates via WimpSim. Capture rate limits can be obtained
by solving \eqref{eq:dm_sol} analytically. As explained in \autoref{sec:dm}, an assumption on
$\langle\sigma_{\rm A}v\rangle$ must be made. For a fair comparison to other searches in the field, the thermal
value $\langle\sigma_{\rm A}v\rangle=3\times10^{-26}\textrm{ cm}^3\textrm{s}$ is assumed. Figure
\ref{fig:cross_sens} shows $\sigma_{\rm SI}$ in function of $\langle\sigma_{\rm A}v\rangle$. The final results for 
$\sigma_{\rm SI}$ can be found by converting capture rates via WimpSim. Figure \ref{fig:xs_sens} illustrates the
90\% C.L. sensitivities for $\sigma_{\rm SI}$ in comparison with other results. For each DM channel and mass
combination, the selection giving the best sensitivity between LE and HE is chosen. The "switch" from LE to HE
happens for every channel at $m_{\chi}\sim100\textrm{ GeV}$. This analysis sensitivities promise competitive limits,
world-leading in the HE selection parameter space.

\clearpage
\section{Conclusions}
\label{sec:conclu}
Sensitivities have been computed for the 8 years search for DM from the center of the Earth. A new likelihood expression has been employed to take into account uncertainties. Adding the energy in 2D distributions with zenith has shown improvements as compared with previous iterations \cite{RenziICRC:2019} of the analysis which were using zenith distributions only. The analysis setting is in the very latest stages, therefore official results are foreseen to be released soon.

\bibliographystyle{ICRC}
\bibliography{references}



\clearpage
\section*{Full Authors List: IceCube Collaboration}




\scriptsize
\noindent
R. Abbasi$^{17}$,
M. Ackermann$^{59}$,
J. Adams$^{18}$,
J. A. Aguilar$^{12}$,
M. Ahlers$^{22}$,
M. Ahrens$^{50}$,
C. Alispach$^{28}$,
A. A. Alves Jr.$^{31}$,
N. M. Amin$^{42}$,
R. An$^{14}$
% ...
\\

\noindent
% ...
$^{12}$ Universit{\'e} Libre de Bruxelles, Science Faculty CP230, B-1050 Brussels, Belgium \\
% ...
$^{14}$ Department of Physics and Laboratory for Particle Physics and Cosmology, Harvard University, Cambridge, MA 02138, USA \\
% ...
$^{17}$ Department of Physics, Loyola University Chicago, Chicago, IL 60660, USA \\
$^{18}$ Dept. of Physics and Astronomy, University of Canterbury, Private Bag 4800, Christchurch, New Zealand \\
% ...
$^{22}$ Niels Bohr Institute, University of Copenhagen, DK-2100 Copenhagen, Denmark \\
% ...
$^{28}$ D{\'e}partement de physique nucl{\'e}aire et corpusculaire, Universit{\'e} de Gen{\`e}ve, CH-1211 Gen{\`e}ve, Switzerland \\
% ...
$^{31}$ Karlsruhe Institute of Technology, Institute for Astroparticle Physics, D-76021 Karlsruhe, Germany  \\
% ...
$^{42}$ Bartol Research Institute and Dept. of Physics and Astronomy, University of Delaware, Newark, DE 19716, USA \\
% ...
$^{50}$ Oskar Klein Centre and Dept. of Physics, Stockholm University, SE-10691 Stockholm, Sweden \\
% ...
$^{59}$ DESY, D-15738 Zeuthen, Germany \\
% ...

\subsection*{Acknowledgements}

\noindent
USA {\textendash} U.S. National Science Foundation-Office of Polar Programs,
U.S. National Science Foundation-Physics Division,
U.S. National Science Foundation-EPSCoR,
Wisconsin Alumni Research Foundation,
Center for High Throughput Computing (CHTC) at the University of Wisconsin{\textendash}Madison,
Open Science Grid (OSG),
Extreme Science and Engineering Discovery Environment (XSEDE),
Frontera computing project at the Texas Advanced Computing Center,
U.S. Department of Energy-National Energy Research Scientific Computing Center,
Particle astrophysics research computing center at the University of Maryland,
Institute for Cyber-Enabled Research at Michigan State University,
and Astroparticle physics computational facility at Marquette University;
Belgium {\textendash} Funds for Scientific Research (FRS-FNRS and FWO),
FWO Odysseus and Big Science programmes,
and Belgian Federal Science Policy Office (Belspo);
Germany {\textendash} Bundesministerium f{\"u}r Bildung und Forschung (BMBF),
Deutsche Forschungsgemeinschaft (DFG),
Helmholtz Alliance for Astroparticle Physics (HAP),
Initiative and Networking Fund of the Helmholtz Association,
Deutsches Elektronen Synchrotron (DESY),
and High Performance Computing cluster of the RWTH Aachen;
Sweden {\textendash} Swedish Research Council,
Swedish Polar Research Secretariat,
Swedish National Infrastructure for Computing (SNIC),
and Knut and Alice Wallenberg Foundation;
Australia {\textendash} Australian Research Council;
Canada {\textendash} Natural Sciences and Engineering Research Council of Canada,
Calcul Qu{\'e}bec, Compute Ontario, Canada Foundation for Innovation, WestGrid, and Compute Canada;
Denmark {\textendash} Villum Fonden and Carlsberg Foundation;
New Zealand {\textendash} Marsden Fund;
Japan {\textendash} Japan Society for Promotion of Science (JSPS)
and Institute for Global Prominent Research (IGPR) of Chiba University;
Korea {\textendash} National Research Foundation of Korea (NRF);
Switzerland {\textendash} Swiss National Science Foundation (SNSF);
United Kingdom {\textendash} Department of Physics, University of Oxford.

\end{document}